\documentclass[12pt]{iopart}

\usepackage{iopams}
\usepackage{citesort}
\usepackage{graphicx}
\graphicspath{{.}{./EPS/}}

\begin{document}

\title{Spin transport and bipolaron density in organic polymers}

\author{P Ingenhoven$^{1,2}$, R Egger$^2$ and U Z\"ulicke$^{1,3}$}

\address{$^1$Institute of Fundamental Sciences and MacDiarmid Institute
for Advanced Materials and Nanotechnology, Massey University (Manawatu
Campus), Private~Bag~11~222, Palmerston North 4442, New Zealand}
\address{$^2$Institut f\"ur Theoretische Physik, Heinrich-Heine-Universit\"at,
D-40225 D\"usseldorf, Germany}
\address{$^3$Centre for Theoretical Chemistry and Physics, Massey
University (Albany Campus), Private Bag 102904, North Shore MSC,
Auckland 0745, New Zealand}
\ead{u.zuelicke@massey.ac.nz}

\begin{abstract}
We present a theory for spin-polarized transport through a generic organic
polymer connected to ferromagnetic leads with arbitrary angle $\theta$
between their magnetization directions, taking into account the polaron and
bipolaron states as effective charge and spin carriers. Within a diffusive
description of polaron-bipolaron transport including polaron-bipolaron
conversion, we find that the bipolaron density depends on the angle $\theta$.
This is remarkable, given the fact that bipolarons are spinless quasiparticles, 
and opens a new way to probe spin accumulation in organic polymers.
\end{abstract}

\submitto{\JPCM}

\maketitle

\section{Introduction}
Recent years have witnessed significant advances in organic electronics,
with interesting new fundamental insights and 
the prospect of new applications and devices functioning at
room temperature \cite{campbell1,kaiser}.  
A particularly interesting aspect comes from the spin degree
of freedom, leading to ``plastic spintronics'' \cite{naber}.  Organic materials
such as polymers may be superior to inorganic semiconductor devices
because of their small spin-orbit and hyperfine couplings, in principle allowing
for very long spin coherence times. 
Moreover, the ease of fabrication and low-temperature
processing of organic materials
is very attractive for possible applications.
Spin transport through $\pi$-conjugated semiconducting organic polymers has 
consequently been studied in a number of recent experiments, 
and evidence for spin-polarized current injection and
giant magnetoresistance in organic spin valves 
\cite{dediu,xiong,pramanik,majumdar} as well as
spin-dependent optical effects \cite{davis,campbell2} have been reported. 

Besides relevance for applications, the unconventional electronic properties of
conducting polymers pose interesting fundamental questions. In undoped
trans-polyacetylene, the charge and spin carriers
are known to be soliton-like excitations, which are characterized by
nontrivial spin-charge relations reflecting electron fractionalization
\cite{ssh}. This raises the possibility of unconventional spin-transport
properties in undoped trans-polyacetylene.  On the other hand, for basically all
 \textit{doped} (nondegenerate) polymers,  it has been established that 
the dominant charge and spin carriers at low energy scales well below
the mean-field Peierls gap $\Delta$ correspond to
\textit{polarons} and \textit{bipolarons} \cite{campbell1,ssh,kirova},
whereas solitons can safely be ignored. As the polaron carries spin 1/2 like
an ordinary electron and the bipolaron is spinless, spin current can only be
carried by the polaron. Nevertheless, as we show below, the bipolaron
density  is affected by spin-polarized transport and can serve as a tool to detect the
latter.

In this work, we discuss spin transport through doped organic polymers, where
polarons and bipolarons are the relevant charge carriers.
In a typical two-terminal geometry (transport along the
$x$ axis), the organic polymer is contacted at $x=0$ and $x=L$ 
by two ferromagnetic (FM) metallic electrodes, where $L$ is the 
length of the polymer.  The left (right) electrode
is characterized by a magnetization unit vector $\hat m_{L}$ ($\hat
m_R$), with the angle $\theta$ between them, 
$\hat m_L\cdot \hat m_R=\cos\theta$.
We do not attempt a microscopic modelling of the interface between a 
FM electrode and the organic polymer, but follow 
the arguments of refs.~\cite{kirova,bussac,basko,xie}, where it has
been established that carriers injected into the polymer tunnel predominantly
into polaron states close to the contact.  
We therefore impose the boundary condition that no bipolaron states
near the boundaries (at $x=0$ and $x=L$) are filled by the injected
current. Both contacts can then be completely
described by spin-dependent conductances $G^\uparrow$ and $G^\downarrow$,
which take into account the spin-dependent density of states in the FM
and the (disorder-averaged) matrix elements for tunneling into
polaron states \cite{brataas1}.
Moreover, for noncollinear magnetizations ($0<\theta<\pi$), one also
has to include the complex-valued mixing conductance
$G^{\uparrow \downarrow}$ reflecting boundary exchange processes 
\cite{brataas2,balents1,balents2}.

Transport in the polymer itself has so far been modelled either
numerically, using lattice simulations of charge transport
\cite{silva,staf1,staf2,schollw}, or analytically, using 
simple master equations \cite{freire}
or drift-diffusion models.  The latter approaches have also
been applied to spin transport \cite{ruden,ren,yu,zhang}. 
Here we use the network theory of ref.~\cite{brataas1,brataas2} 
combined with a diffusive model 
to obtain spin-transport properties of a doped organic
polymer sandwiched between two FM electrodes with noncollinear
magnetization directions (arbitrary $\theta$).  In the absence of 
bipolarons and for very high temperatures, this problem 
has been studied in ref.~\cite{yu}.  Here we present a generalization
including the polaron-bipolaron conversion process, 
and also study the low-temperature quantum-degenerate limit.

\section{Model}
The energy-dependent polaron distribution function $\hat f_P(x,\epsilon)$
at location $0<x<L$ can be decomposed into a spin-independent 
scalar part $f_0(x,\epsilon)$ and a spin-polarization
 vector ${\boldsymbol f}(x,\epsilon)$,
\begin{equation}
\hat f_P (x,\epsilon) = f_0 \sigma_0 + {\boldsymbol f}\cdot {\boldsymbol \sigma},
\end{equation}
with the standard Pauli matrices $\sigma_i$ in spin space; 
$\sigma_0$ is the unit matrix, and we assume homogeneity in the 
transverse direction.  Note that a polaron has charge $e$ and spin $1/2$.
Another important charge carrier in organic polymers
is the spinless bipolaron, with charge $2e$ and the 
scalar distribution function $f_{BP}(x,\epsilon)$ \cite{campbell1,ssh}.
With the average density of states  $\rho(\epsilon)$, 
we introduce normalized densities by integrating the
distribution functions over energy,
\begin{eqnarray}\nonumber
\hat n_P (x) & =& \int d\epsilon \rho(\epsilon) \hat f_P(x,\epsilon)
= n_0(x)+ {\boldsymbol n}(x) \cdot {\boldsymbol\sigma}, \\ \label{dens}
n_{BP}(x) &=&  \int d\epsilon \rho(\epsilon) f_{BP}(x,\epsilon).
\end{eqnarray}
These densities are defined relative to an equilibrium reference
value, and reflect nonequilibrium charge and spin accumulation
in the polymer.  Since our model does not include the 
quasiparticle states outside the 
mean-field gap $\Delta$, but only retains the polaron and
bipolaron states inside the gap, we choose 
$\rho(\epsilon)\simeq \Delta^{-1}\Theta(\Delta-|\epsilon|)$.

In typical organic polymers, disorder is present and implies 
diffusive transport for both polarons and bipolarons, with 
the respective diffusion constants $D_P$ and $D_{BP}$.
The equations of motion for $\hat n_P(x,t)$ and  $n_{BP}(x,t)$ are thus
\begin{eqnarray}\label{dtnp}
\partial_t \hat n_P &=& D_P \partial_x^2 \hat n_P - \tau_{sf}^{-1}
\left( \hat n_P- n_0 \sigma_0\right) + i [{\boldsymbol h}\cdot{\boldsymbol \sigma}, \hat n_P]_-  - S_P \sigma_0,   \\  
\label{dtnbp} \partial_t n_{BP} &=& D_{BP} \partial_x^2 n_{BP} + S_{P},
\end{eqnarray}
where $\tau_{sf}$ is the polaron spin-relaxation time
and $S_P$ models conversion processes between polarons and bipolarons
\cite{zhang},
\begin{equation}\label{sp}
S_P (x) =  k \left (n_0^2 -{\boldsymbol n}^2 \right)  - b n_{BP}.
\end{equation}
The parameter $k$ describes the local recombination rate for two 
polarons of opposite spin forming a bipolaron, while $b$ comes
from the reverse process, where a bipolaron decomposes into
two polarons of opposite spin. 
The spin-precession term in (\ref{dtnp}) comes from an applied
homogeneous magnetic field, where ${\boldsymbol h}=g\mu_B {\boldsymbol B}/\hbar$. 
We are interested in the steady-state case, where 
$\partial_t \hat n_P=\partial_t n_{BP}=0$ in (\ref{dtnp})
and (\ref{dtnbp}).  According to Fick's law,
the stationary spin-dependent particle current in the polymer is then encoded  
in the $2\times 2$ matrix (in spin space)
\begin{equation}\label{current}
\hat J (x) = - D_P \partial_x \hat n_P(x) -D_{BP} \partial_x n_{BP}(x) \sigma_0.
\end{equation}
Equation (\ref{dtnp}) yields a decoupled equation for the
spin polarization vector,
\begin{equation}\label{d2nvec}
D_P \partial_x^2 {\boldsymbol n}(x)=  
\left( \begin{array}{ccc} \tau_{sf}^{-1} & -h_z & h_y \\
h_z & \tau_{sf}^{-1} & -h_x \\ 
-h_y & h_x & \tau_{sf}^{-1}  \end{array} \right) 
 \cdot {\boldsymbol n}(x).
\end{equation}
Given the solution to (\ref{d2nvec}),
by taking the scalar part of (\ref{dtnp}) and combining
it with (\ref{dtnbp}), the bipolaron density is determined by 
\begin{equation}\label{bipol}
n_{BP}(x) = -\frac{D_P}{D_{BP}} \left( {\cal O}\frac{x}{L}
 + {\cal P}+n_0(x) \right),
\end{equation}
with two integration constants ${\cal O}$ and ${\cal P}$.
The only nontrivial equation that needs to be solved is given by 
\begin{equation}\label{nontriv}
D_P \partial_x^2 n_0 = k(n_0^2-{\boldsymbol n}^2 ) + 
\frac{bD_P}{D_{BP}} \left( \frac{{\cal O}}{L} x +{\cal P}+n_0\right ).
\end{equation}
As discussed above, we impose the boundary condition
\begin{equation}\label{nobip}
n_{BP}(0)=n_{BP}(L)=0,
\end{equation}
since tunneling into the polymer involves only polaron states.
With  (\ref{bipol}), this
implies boundary conditions for  (\ref{nontriv}),
\begin{equation}\label{bc1}
n_{0}(0)= -{\cal P},\quad n_0(L)= -({\cal P}+{\cal O}).
\end{equation}
In order to solve  (\ref{d2nvec}), we need six additional
integration constants. We therefore have to
specify boundary conditions reflecting spin and charge
current continuity at the contacts to the left and right FMs. 
The FMs are taken as reservoirs with identical 
temperature $T$ and chemical potentials $\mu_{L/R}$, where the 
applied voltage is $eV=\mu_L-\mu_R$. As before, we 
introduce (normalized) densities,
\begin{equation}
n^{FM}_{L/R}= \int d\epsilon \rho(\epsilon) n_F(\epsilon-\mu_{L/R}),
\end{equation}
with the Fermi function $n_F(\epsilon)=1/[e^{\epsilon/k_B T}+1]$.
Boundary conditions then follow by relating the
current (\ref{current}) at $x=0$ ($x=L$) to the
injected current at the left (right) interface \cite{brataas2},
\begin{eqnarray}\label{inj0}
\hat J(0) &=&- \sum_{\sigma=\uparrow,\downarrow} G^\sigma 
\hat u_L^\sigma (\hat n_P(0)-n_L^{FM} \sigma_0) \hat u_L^\sigma -
\left(G^{\uparrow \downarrow} \hat u^\uparrow_L \hat n_P(0) \hat
u^\downarrow_L + {\rm h.c.}\right) , \\  \label{injL}
\hat J(L) &=& \sum_{\sigma=\uparrow,\downarrow} G^\sigma
\hat u_R^\sigma (\hat n_P(L)-n_R^{FM} \sigma_0) \hat u_R^\sigma 
+ \left(G^{\uparrow \downarrow} \hat u^\uparrow_R \hat n_P(L) \hat
u^\downarrow_R + {\rm h.c.}\right) .
\end{eqnarray}
Note that  (\ref{nobip}) implies that bipolarons do not
enter this boundary condition.
The matrices $\hat u^\sigma_{L,R}=\frac12(1+\sigma \hat m_{L,R}\cdot {\boldsymbol
\sigma})$ project the spin direction $\sigma=\uparrow,\downarrow=+,-$ 
in the polymer onto the respective FM magnetization direction.
For simplicity, we assumed identical spin-polarized ($G^\uparrow,
G^\downarrow$) and mixing ($G^{\uparrow\downarrow}$) conductances
for both contacts. They must obey ${\rm Re} 
G^{\uparrow\downarrow}\ge (G^\uparrow+G^\downarrow)/2$ \cite{brataas2}.
The $2\times 2$ matrix equations (\ref{inj0}) and (\ref{injL}) allow
to determine the eight integration constants, and thereby
yield the spin-polarized current through the system
for arbitrary $\theta$. Moreover, this gives access to the 
bipolaron density from  (\ref{bipol}) after solving  (\ref{nontriv}).
We stress that none of the eight integration constants 
depends on the parameters $k$ and $b$ in  (\ref{sp}).

{}From  (\ref{current}) and (\ref{bipol}), we can immediately 
see that charge current $J_c=D_P {\cal O}/L$ is conserved, 
\begin{equation}
\hat J(x)= J_c \sigma_0 + {\boldsymbol J}_s(x) \cdot {\boldsymbol \sigma},
\end{equation}
and the spin current,  ${\boldsymbol J}_s(x) = -D_P \partial_x {\boldsymbol n}(x)$,
follows from the solution of  (\ref{d2nvec}). 
Remarkably, both $J_c$ and ${\boldsymbol J}_s(x)$ are independent of the 
polaron-bipolaron transition rates $k$ and $b$ in  (\ref{sp}),
and the spin-dependent current alone cannot detect the presence
of bipolarons in the polymer. Nevertheless, as we show below, the
bipolaron density $n_{BP}(x)$, which is induced by the 
nonequilibrium spin accumulation in the polymer, is sensitive
to these rates.  As a useful measure, we will employ the integrated 
density,
\begin{equation}\label{ntheta}
N(\theta) = \int_0^L dx\ n_{BP}(x;\theta).
\end{equation}
The $\theta$-dependence of the bipolaron density is then 
encoded in the dimensionless quantity
\begin{equation}\label{rdef}
R(\theta) = \frac{N(0)-N(\theta)}{N(0)-N(\pi)}.
\end{equation}
By definition, this quantity interpolates between $R(0)=0$ and 
$R(\pi)=1$ as $\theta$  is varied from the parallel to the
antiparallel configuration.

\section{Collinear case: a readily solvable limit}
We first discuss a simple yet important limit, where a direct
analytical solution can be obtained.  This limit is defined by 
collinear magnetizations, $\hat m_R= p \hat m_L$ with $p=\pm$ (parallel
or antiparallel configuration) and $\hat m_L=\hat e_z$. 
Moreover, we consider the length of
the polymer as short compared to the spin coherence length,
$L\ll \sqrt{D_P\tau_{sf}}$, and put ${\boldsymbol h}=0$ (no magnetic field).
In that case,  (\ref{d2nvec}) has the
general solution ${\boldsymbol n}(x)=-( {\boldsymbol F} x/L + {\boldsymbol G})$, with 
constant vectors ${\boldsymbol F}$ and ${\boldsymbol G}$.  
For $\hat m_L=\hat e_z=\pm \hat m_R$, the boundary conditions
(\ref{inj0}) and (\ref{injL}) imply that the $x$ and $y$ components
of both vectors vanish, and the spin current is conserved,
\begin{equation}\label{spincur}
{\boldsymbol n}(x) = -\hat e_z \left( \frac{{\cal F}}{L} x+{\cal G}\right),
\quad {\boldsymbol J}_s= \frac{D_P {\cal F} }{L} \hat e_z.
\end{equation}
The four remaining integration constants $({\cal O},{\cal P},
{\cal F},{\cal G})$ readily follow by solving the 
boundary conditions (\ref{inj0}) and (\ref{injL}) \cite{brataas2}. For
the parallel ($p=+$) configuration, they are 
\begin{eqnarray}\nonumber 
{\cal O}_+ &=& 2({\cal P}_+ +\bar\mu) =
\frac{G^\uparrow G^\downarrow + 2(G^\uparrow+G^\downarrow) G_P}
{(G^\uparrow+2G_P)(G^\downarrow +2G_P)}  eV, \\  
\label{coeffpar}
{\cal F}_+ &=& -2{\cal G}_+ =  \frac{(G^\uparrow-G^\downarrow )G_P}
{(G^\uparrow+2G_P)(G^\downarrow +2G_P)}  eV, 
\end{eqnarray}
while for the antiparallel case, we find
${\cal F}_- =0$ and
\begin{eqnarray}
\nonumber
{\cal O}_- &=& -2({\cal P}_-+\bar\mu)= \frac{ G^\uparrow G^\downarrow  }
{G^\uparrow G^\downarrow+ 2 G_P(G^\uparrow +G^\downarrow) } eV,
 \\ {\cal G}_- &=& - \frac{ (G^\uparrow-G^\downarrow)G_P } 
{G^\uparrow G^\downarrow+ 2 G_P(G^\uparrow +G^\downarrow) } \frac{eV}{2},
\label{coeffanti}
\end{eqnarray}
where $\bar \mu=(\mu_L+\mu_R)/2$ is the mean chemical potential
and $G_P\equiv D_P/L$.
The charge current for the respective configuration
is then $J_c=G_P {\cal O}_\pm$, while the spin current is
${\boldsymbol J}_s=G_P {\cal F}_\pm \hat e_z$.  

\begin{figure}[t]
\begin{indented}
\item[] \includegraphics[width=0.5\textwidth]{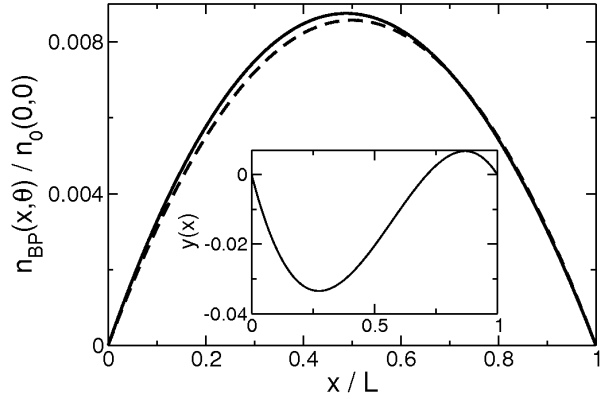}
\caption{\label{f1}Bipolaron density $n_{BP}(x,\theta)$ for collinear 
magnetizations, i.e., $\theta=0$ (solid curve) and $\theta=\pi$ (dashed curve),
as obtained from (\ref{nbp}).   We use a representative parameter set 
for hole transport \cite{yu}. In units where $G_P=D_P/L=1$, the
parameters are $D_{BP}=2L/3, G^\uparrow=10^4, G^\downarrow=10^{-2}, 
k=D_P/10, b=D_{BP}/10$, $\Delta=3.5$, $\bar \mu=-2$, $T=0$,
and $eV=1$. The inset shows $y(x)=[n^{(-)}_{BP}(x)-n_{BP}^{(+)}(x)]
/n_{BP}^{(-)}(L/2)$ for the same curves.}
\end{indented}
\end{figure}

The remaining task is to solve (for given $p=\pm$) the nonlinear equation
(\ref{nontriv}) for $n_0(x)$ under the boundary condition (\ref{bc1}),
using  (\ref{spincur}), (\ref{coeffpar}) and (\ref{coeffanti}).
 Since the transition
rates $k$ and $b$ are known to be small \cite{zhang}, we  
use a perturbative iteration scheme and write 
\begin{equation}\label{ansatz}
n_0(x)=-\frac{{\cal O}}{L} x - {\cal P} + \tilde n_0 (x).
\end{equation}
For $k=b=0$, this Ansatz solves  (\ref{nontriv}) under the
correct boundary conditions when putting $\tilde n_0(x)=0$;
note that the bipolaron density is directly proportional to $\tilde n_0(x)$,
see  (\ref{bipol}).
For small but finite $k,b$, we then insert  (\ref{ansatz})
into  (\ref{nontriv}) and linearize in $\tilde n_0$.  
This yields a second-order differential equation for $\tilde n_0(x)$,
which needs to be solved under Dirichlet boundary conditions at $x=0$ and
$x=L$.  The solution gives the bipolaron density for the parallel
and antiparallel configuration in closed form,
\begin{eqnarray}\label{nbp}
n^{(p=\pm)}_{BP}(x) &=& -\frac{k}{D_{BP}} \Biggl({\cal C}_\pm x+ 
({\cal P}_\pm^2-{\cal G}_\pm^2 )\frac{x^2}{2} + \frac{{\cal P}_\pm
{\cal O}_\pm-{\cal F}_\pm {\cal G}_\pm}{L}\frac{x^3}{3} \nonumber \\
&& \hspace{4cm} + \frac{{\cal O}_\pm^2-{\cal F}_\pm^2}{L^2}
\frac{x^4}{12}\Biggr),
\end{eqnarray}
where the integration constant ${\cal C}_\pm$ follows from the condition
$n_{BP}(L)=0$.  The integrated bipolaron density (\ref{ntheta}) is then
given by 
\begin{equation}\label{intf}
N_{\pm} = \frac{kL^3}{12 D_{BP}} \Biggl( {\cal P}_\pm^2-{\cal G}_\pm^2
+{\cal P}_\pm {\cal O}_\pm -{\cal F}_\pm {\cal G}_\pm + \frac{3}{10}
({\cal O}_\pm^2-{\cal F}_\pm^2)\Biggr).
\end{equation}
Note that $N_+\ne N_-$ follows immediately from  (\ref{coeffpar})
and (\ref{coeffanti}), indicating that the bipolaron density
indeed is sensitive to the spin accumulation in the polymer.
The bipolaron density (\ref{nbp}) is shown in \fref{f1},
taking parameters for sexithienyl as organic spacer \cite{yu}.
One clearly observes a difference between the parallel and the
antiparallel configuration.  Although the current is not sensitive
to the polaron-bipolaron transition rates $k$ and $b$, the 
bipolaron density is influenced by the nonequilibrium spin accumulation
in the polymer.

\section{Noncollinear magnetization}
\begin{figure}[t]
\begin{indented}
\item[] \includegraphics[width=0.5\textwidth]{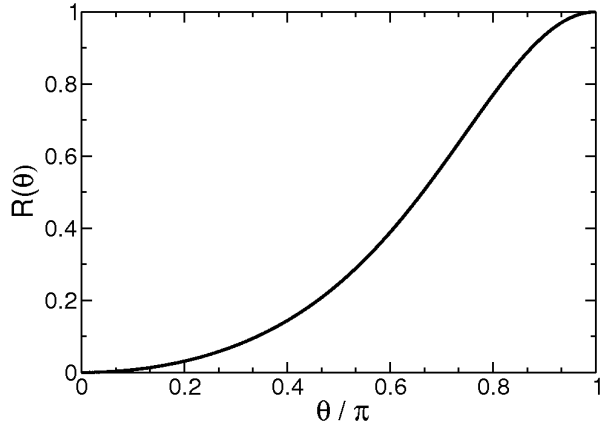}
\caption{\label{f2}Spin-accumulation sensitivity $R(\theta)$, 
see  (\ref{rdef}), of the bipolaron density as a function
of the magnetization tilt angle.  Parameters are as in \fref{f1}, additionally we set $\mbox{Re}G^{\uparrow\downarrow}=\mbox{Im}G^{\uparrow\downarrow}=5.1\times 10^3$.}
\end{indented}
\end{figure}

In the general case of arbitrary angle $\theta$ between $\hat m_L$ and
$\hat m_R$, one can solve the problem
in an analogous manner but the equations become less transparent. The
main difference is that now the mixing conductance $G^{\uparrow\downarrow}$
 has to be  taken into account.  However, as reported previously
\cite{yu}, we find that the results are practically independent
of the precise choice for $G^{\uparrow\downarrow}$. 
We find a smooth crossover between the limiting values for $\theta=0$ and
$\theta=\pi$, see  (\ref{nbp}), 
illustrated for the integrated bipolaron density (\ref{ntheta}) 
in \fref{f2}.  

\section{Conclusions}

In this work, we have discussed spin transport in doped organic polymers,
employing a diffusive description of polaron and bipolaron 
transport. In a two-terminal setup, where the polymer is
sandwiched by (generally noncollinear) ferromagnetic
electrodes, the problem can be solved analytically by exploiting the smallness
of the polaron-bipolaron transition rates $k$ and $b$. 
While the spin-dependent current through the device turns out to 
be independent of $k$ and $b$, the nonequilibrium bipolaron density
is a sensitive probe of spin accumulation. The possibility to measure this
density in optical-absorption experiments \cite{patil,salaneck,blythe}, e.g., by adapting
charge-modulation techniques \cite{harrison} to the two-terminal transport
geometry considered here, may offer a novel way to probe spin accumulation
in organic polymers. Such an optical method would be complementary to the
usual magnetoresistance measurement of spin accumulation and 
could thus serve as another means to independently verify spin-injection 
efficiencies in organic polymers~\cite{vardeny}.

Our work generalizes previous studies where bipolarons were neglected \cite{yu}
or only a single ferromagnet-polymer interface was considered \cite{zhang}. We also
treat the nonequilibrium situation due to an applied voltage self-consistently instead of
postulating the existence of a uniform electric field \cite{zhang}.
We mention in passing that results from a recent Monte Carlo simulation
\cite{bobbert} have elucidated the importance of bipolaronic effects for a nontraditional type of magnetoresistance that occurs in conducting polymers
in the absence of magnetic contacts.
Another recent theoretical study \cite{ren2} on magnetoresistance in polymers with 
polaron and bipolaron carriers used a diffusive approach and magnetic contacts
(FM-polymer-FM configuration). In contrast to our work, however, ref.~\cite{ren2} does
not take into account conversion processes between polaron and bipolaron states,
but simply assumes a constant density of bipolarons and includes this into the 
transport calculations. Surprisingly, a dependence of the magnetoresistance 
 on the ratio of bipolarons and polarons is reported~\cite{ren2}, 
whereas we find the spin-polarized current to be independent of the bipolaron
formation rate. Our finding can be traced back to the well-established \cite{basko,xie}
suppression of tunneling into bipolaron states near the interface with a FM 
electrode.  This feature is ignored when simply asssuming a constant 
bipolaron density.

\ack
P.I.\ is supported by a Massey University Doctoral Scholarship.
Additional funding from the ESF network INSTANS is gratefully acknowledged.
U.Z.\ thanks A B Kaiser (Victoria University of Wellington) for useful
discussions.

\end{document}